# 发电机动态状态估计的一种虚假数据注入攻击方法

李扬[1]  李智[1]  陈亮[2]  李国庆[1]

（1 东北电力大学电气工程学院，吉林省 吉林市 132012；
2 南京信息工程大学自动化学院，江苏省 南京市 210044）

**摘要** 精确、可靠的发电机动态状态量对电力系统的实时监测和控制至关重要。信息攻击的出现给发电机状态估计带来了新的挑战。其中，虚假数据注入（false data injection，FDI）攻击通过对量测装置注入虚假数据，恶化了状态估计的精度。为此，本文首次提出了一种针对发电机动态状态估计的 FDI 攻击模型。首先，采用泰勒公式将发电机量测方程线性化；其次，根据 FDI 攻击前后量测残差不变的原理，建立了攻击向量的表达式，将其施加在量测量中，从而躲避常规的不良数据检测、成功实施 FDI 攻击；然后，根据攻击程度分别设定了三种攻击情形，通过容积卡尔曼滤波(cubature Kalman filter，CKF)和抗差容积卡尔曼滤波(robust CKF，RCKF)对所提的三种不同情形的 FDI 攻击进行验证；最后，IEEE 9 节点系统和新英格兰 16 机 68 节点系统的仿真结果验证了所提 FDI 攻击的有效性。

**关键词**：动态状态估计 信息攻击 发电机 容积卡尔曼滤波 虚假数据注入攻击 PMU 数据

**中图分类号**：TM71

## A false data injection attack method for generator dynamic state estimation

*Li Yang [1], Li Zhi [1], Chen Liang [2], Li Guoqing [1]*
(1 School of Electrical Engineering, Northeast Electric Power University, Jilin 132012, Jilin Province, China;
2 School of Automation, Nanjing University of Information Science & Technology, Nanjing 210044, Jiangsu Province, China)

**Abstract** Accurate and reliable dynamic state quantities of generators are very important for real-time monitoring and control of the power system. The emergence of cyber attacks has brought new challenges to the state estimation of generators. Especially, false data injection (FDI) attacks deteriorate the accuracy of state estimation by injecting the false data into the measurement device. In this regard, this paper proposes for the first time an FDI attack model based on the dynamic state estimation of generators. Firstly, Taylor's formula was used to linearize the generator's measurement equation. Secondly, according to the principle that the measurement residuals before and after the FDI attack are equal, the expressions of the attack vectors were established, and they were applied to the measurement quantities to avoid the conventional bad data detection. Thereby, the FDI attacks were successfully implemented. Then, three attack scenarios were set according to the degree of the FDI attacks, and they were tested by the cubature Kalman filter (CKF) and the robust cubature Kalman filter (RCKF). Finally, the simulation results of the IEEE 9-bus system and the New England 16-machine 68-bus system verify the effectiveness of the proposed FDI attacks.

**Keywords**: Dynamic state estimation, cyber attacks, generator, cubature Kalman filter, false data injection attacks, PMU data

## 0 引言

精确、可靠的发电机动态状态量对电力系统的实时监测和控制至关重要[1-3]。作为一个典型的信息物理系统，电力系统已经出现并运行了一百多年，但是日益增多的信息攻击、自然灾害和对通信控制的依赖导致了新型故障源的产生和连锁故障的发生。同时，随着可再生能源的大量接入，电力系统运行的不确定性也随之增大[4]。所有这些转变都给维持系统安全、可靠运行带来了新的挑战。同步相量测量单元（phasor measurement unit，PMU）作为电力系统状态估计与动态监视的重要部分，在电力系统安全稳定评估和监控中发挥了重要作用[5-6]。

当前电力系统的运行面临着严峻的信息攻击威胁，其中虚假数据注入（false data injection，FDI）攻击就是信息攻击的一种典型代表。该攻击对电力系统中量测装置产生的量测数据进行虚假数据注入，且具有避开不良数据检测的能力，使估计结果较正常结果偏差较大。从而影响电力系统在线安全评估，使安全约束经济调度方案偏离最优解，增加系统运行成本，甚至造成大范围甩负荷的事件发生。2015年 12 月 23 日，乌克兰国家电网突发停电事故就是FDI 攻击电力系统的现实案例[7]。

近年来，国内外对 FDI 攻击和发电机状态估计的研究取得了一定进展。就 FDI 攻击而言，针对攻击的建模、检测及防御已有大量文献对此进行了研究。其中，文献[8]通过引入攻击向量的松弛误差建立了一种 FDI 攻击模型；文献[9]使用主成分分析法对 FDI 攻击进行建模,并验证了所提攻击的隐蔽性；文献[10，11]分别基于网络拓扑结构建立了针对直流潮流和交流潮流的 FDI 攻击模型；而文献[12]又将直流模型扩展为更一般的线性模型，推导出了基于该模型的 FDI 攻击；文献[13]则依靠量测矩阵的低秩特征和攻击矩阵的稀疏性对 FDI 攻击进行检测；文献[14]提出了一种动态估计的风险缓解策略以消除 FDI 攻击的威胁；文献[15]研究了在 FDI 攻击下卡尔曼滤波器对电力系统动态估计的影响。在发电机动态状态估计领域，各文献也使用了很多非线性滤波的方法来解决此问题。比如：文献[1，16，17]就分别采用容积卡尔曼滤波（cubature Kalman filter，CKF）、无迹卡尔曼滤波（unscented Kalman filter，UKF）和扩展卡尔曼滤波（extended Kalman filter，EKF）的方法进行发电机动态状态估计；其中 EKF 通过泰勒公式展开将非线性问题线性化，但由于忽略了泰勒展开式中的高次项，进而对估计精度产生了一定的影响[18]；UKF 利用无迹变换来解决非线性问题，无需进行雅克比公式的计算，但其估计精度很大程度上取决于 sigma 点的分布情况和状态维数，这也在一定程度上限制了 UKF 的应用[19]；CKF 则采用球面-径向准则生成一组等权值的容积点，进而解决了发电机动态状态估计中的非线性滤波问题，其在处理估计偏差和状态维度的问题上优于 EKF 和 UKF。综上，在状态估计领域内，多数文献研究了基于电力系统稳态潮流模型的 FDI 攻击，迄今尚未见有在发电机动态状态估计中施加 FDI 攻击的相关研究报道。

本文提出了一种新的针对发电机动态状态估计的虚假数据注入攻击方法。首先，采用泰勒公式将发电机量测方程线性化；其次，根据 FDI 攻击前后量测残差不变的原理，建立了攻击向量的表达式，将其施加在量测量中，从而躲避常规的不良数据检测、成功实施 FDI 攻击；然后，根据攻击程度设定了三种攻击情形，通过容积卡尔曼滤波和抗差容积卡尔曼滤波(robust CKF，RCKF)对所提的不同情形的 FDI 攻击进行验证；最后，IEEE 9 节点系统和新英格兰 16 机 68 节点系统的仿真结果验证了所提 FDI 攻击的有效性。

# 1 发电机状态估计建模

## 1.1 非线性动态系统的数学模型

建立非线性动态系统数学模型是对发电机动态状态估计研究的前提，其状态方程和量测方程表示如下[1]

$$\begin{cases} x_{k+1} = f(x_k, u_k, v_k) \\ z_{k+1} = h(x_{k+1}, u_{k+1}, w_{k+1}) \end{cases} \quad (1)$$

上式中，$x$，$u$，$z$ 分别表示状态向量，控制向量和量测向量，$v$ 和 $w$ 分别为系统噪声向量和量测噪声向量，其分别服从均值为 0、误差方差阵为 $Q$ 和 $R$ 的正态分布，其中 $Q$ 为系统噪声方差矩阵，$R$ 为量测噪声方差矩阵；$k$ 为时刻。

## 1.2 发电机模型

发电机动态状态估计的状态方程如下[20,21]：

$$\begin{cases} \dot{\delta} = \omega - 1 \\ \dot{\omega} = \dfrac{1}{T_J}\left[T_m - T_e - D(\omega - 1)\right] \\ \dot{E}_q' = \dfrac{1}{T_{d0}'}\left[E_f - E_q' - (X_d - X_d')i_d\right] \\ \dot{E}_d' = \dfrac{1}{T_{q0}'}\left[-E_d' + (X_q - X_q')i_q\right] \end{cases} \quad (2)$$

$$\begin{cases} i_d = \dfrac{E_q' - U\cos(\delta - \varphi)}{X_d'} \\ i_q = \dfrac{U\sin(\delta - \varphi) - E_d'}{X_q'} \end{cases} \quad (3)$$

式中：$E_d'$ 表示发电机 $d$ 轴暂态电动势；$E_q'$ 表示发电机 $q$ 轴暂态电动势；$\omega$ 表示电角速度标幺值；惯性时间常数用 $T_J$ 表示；$T_m$ 表示发电机机械转矩；励磁电动势用 $E_f$ 表示；$T_e$ 表示发电机电磁转矩；$D$ 表示阻尼系数；$T_{d0}'$ 表示发电机 $d$ 轴暂态时间常数；$X_d$ 表示发电机 $d$ 轴电抗；$X_d'$ 表示发电机 $d$ 轴暂态电抗；$i_d$ 表示发电机 $d$ 轴输出电流分量；$T_{q0}'$ 表示发电机 $q$ 轴暂态时间常数；$X_q$ 表示发电机 $q$ 轴电抗；$X_q'$ 表示 $q$ 轴暂态电抗；$i_q$ 表示发电机 $q$ 轴输出电流分量；$\delta$ 表示功角；$U$ 和 $\varphi$ 分别表示发电机出口电压幅值和相角。

量测变量包括 $\delta$、$\omega$ 和 $P_e$，量测方程如下[3]：

$$\begin{cases} \delta^z = \delta \\ \omega^z = \omega \\ P_e^z = \dfrac{U^2}{2}\sin(2\delta - 2\varphi)\left(\dfrac{1}{X_q'} - \dfrac{1}{X_d'}\right) \\ \qquad + \dfrac{U\sin(\delta - \varphi)E_q'}{X_d'} + \dfrac{U\cos(\delta - \varphi)E_d'}{X_q'} \end{cases} \quad (4)$$

式中，$\omega^z$ 和 $\delta^z$ 分别为发电机角速度和功角的量测量，$P_e^z$ 为发电机电磁功率量测量。

发电机量测误差方差矩阵 $R_{k+1}$ 表示为：

$$R_{k+1} = \begin{bmatrix} \sigma_{\delta_z}^2 & 0 & 0 \\ 0 & \sigma_{\omega_z}^2 & 0 \\ 0 & 0 & \sigma_{p_{ez}}^2 \end{bmatrix} \quad (5)$$

式中 $\sigma_{\delta_z}^2$，$\sigma_{\omega_z}^2$，$\sigma_{p_{ez}}^2$ 分别表示功角量测方差、角速度量测方差和电磁功率量测方差，$\sigma_{\delta_z}^2 = 4°$，$\sigma_{\omega_z}^2 = 1 \times 10^{-6}$。

$$P_{e_z} = \left(\frac{\partial P_e}{\partial U}\right)^2 \sigma_U^2 + \left(\frac{\partial P_e}{\partial \varphi}\right)^2 \sigma_\varphi^2 \quad (6)$$

在上式中，$\sigma_U = 0.2\%$，$\sigma_\varphi = 0.2°$。

在发电机模型中，状态量、量测量和控制量分别表示如下：

$$\begin{cases} x = [\delta, \omega, E_q', E_d'] \\ z = [\delta^z, \omega^z, P_e^z] \\ u = [T_m, E_f, U, \varphi] \end{cases} \quad (7)$$

## 2 虚假数据注入攻击模型

在发电机动态状态估计中，状态变量统一表示为 $x = [x_1, x_2, x_3, \ldots x_k]^T$，量测量统一表示为 $z = [z_1, z_2, z_3, \ldots z_l]^T$。量测方程如下所示：

$$z = h(x) + e \quad (8)$$

其中

$$h(x) = [h_1(x_1, \ldots, x_k), \\ h_2(x_1, \ldots, x_k), \\ \ldots \\ h_l(x_1, \ldots, x_k)]^T \quad (9)$$

表示状态量和量测量之间的非线性关系，$e = [e_1, e_2, \ldots, e_l]^T$ 表示量测误差。

以 $h_1$ 函数为例，使用泰勒公式对 $h_1$ 函数进行扩展，由于发电机各状态量参数在机电暂态过程中不能突变，$h_1$ 函数中的高次项可以被忽略，因此能够对发电机量测方程线性化，具体如下所示：

$$\begin{aligned} z_1 &= h_1(x_1, \ldots, x_k) + e_1 \\ &= h_1(x_1^{(0)}, \ldots, x_k^{(0)}) + \left.\frac{\partial h_1}{\partial x_1}\right|_0 (x_1 - x_1^{(0)}) \\ &\quad + \ldots + \left.\frac{\partial h_1}{\partial x_k}\right|_0 (x_k - x_k^{(0)}) + e_1 \\ &= h_1(x_1^{(0)}, \ldots, x_k^{(0)}) + \left(\left.\frac{\partial h_1}{\partial x_1}\right|_0 x_1 - \left.\frac{\partial h_1}{\partial x_1}\right|_0 x_1^{(0)}\right) \\ &\quad + \ldots + \left(\left.\frac{\partial h_1}{\partial x_k}\right|_0 x_k - \left.\frac{\partial h_1}{\partial x_k}\right|_0 x_k^{(0)}\right) + e_1 \\ &= \left.\frac{\partial h_1}{\partial x_1}\right|_0 x_1 + \ldots + \left.\frac{\partial h_1}{\partial x_k}\right|_0 x_k + \bar{e}_1 \end{aligned} \quad (10)$$

上式中，$x_1^{(0)}, \ldots x_k^{(0)}$ 表示接近状态量 $x_1, \ldots x_k$ 的设置初始值；$\left.\frac{\partial h_1}{\partial x_1}\right|_0, \ldots, \left.\frac{\partial h_1}{\partial x_k}\right|_0$ 表示将 $x_1^{(0)}, \ldots, x_k^{(0)}$ 代入偏导数公式后的计算结果；$\bar{e}_1$ 为常数矩阵。因此，公式（8）可进一步表示为：

$$z = Hx + \bar{e} \quad (11)$$

$$H = \begin{bmatrix} \frac{\partial h_1}{\partial x_1} & \frac{\partial h_1}{\partial x_2} & \frac{\partial h_1}{\partial x_3} & \cdots & \frac{\partial h_1}{\partial x_k} \\ \frac{\partial h_2}{\partial x_1} & \frac{\partial h_2}{\partial x_2} & \frac{\partial h_2}{\partial x_3} & \cdots & \frac{\partial h_2}{\partial x_k} \\ & & \cdots & & \\ \frac{\partial h_l}{\partial x_1} & \frac{\partial h_l}{\partial x_2} & \frac{\partial h_l}{\partial x_3} & \cdots & \frac{\partial h_l}{\partial x_k} \end{bmatrix} \quad (12)$$

上式中，$H$ 为雅克比矩阵，$\bar{e}$ 可被视为新的量测误差。将式（4）代入式（12），可得 $H$ 的具体表达式：

$$H = \begin{bmatrix} 1 & 0 & 0 & 0 \\ 0 & 1 & 0 & 0 \\ L_1 & 0 & L_2 & L_3 \end{bmatrix} \quad (13)$$

式中，$L_1 = \frac{\partial P_e^z}{\partial \delta}$，$L_2 = \frac{\partial P_e^z}{\partial E_q'}$，$L_3 = \frac{\partial P_e^z}{\partial E_d'}$。

在发电机实际运行中，不良数据时常出现，严重影响电力系统状态估计精度[19]。残差方程的应用能够很大程度上消除了不良数据的影响。因此，文中把基于残差的检测法应用于对量测数据的检测。

定义残差方程如下式所示[22]：

$$r = z - \hat{z} \quad (14)$$

式中 $\hat{z} = h(\hat{x})$ 表示量测量 $z$ 的估计值。

依据残差的定义并结合式（8）–（13），上式可改写如下：

$$r = z - H\hat{x} \quad (15)$$

若用 $a = [a_1, \ldots, a_l]^T$ 表示攻击者在量测值中注入的虚假数据向量，则实际的量测数据为 $z_a = z + a$，$c = [c_1, \ldots, c_k]^T$ 表示状态变量中由 FDI 攻击而带入的误差向量，其可服从均值为零，不同标准差的高斯分布[23-24]，进而状态变量估计值变为 $\hat{x}_a = \hat{x} + c$。因此，FDI 攻击后的量测残差具体表示如下[9]：

$$\begin{aligned} \|r\| &= \|z_a - H\hat{x}_a\| \\ &= \|z + a - H(\hat{x} + c)\| \\ &= \|z - H\hat{x} + a - Hc\| \end{aligned} \quad (16)$$

显然，当 $a = Hc$，有式（17）成立，结果显示在 FDI 攻击前后的量测残差相等，导致在以残差方程为根据的基础上，不良数据检测无法识别虚假数据，因此量测数据被成功引入 FDI 攻击，攻击下的量测数据比真实值偏差较大，将直接破坏电力系统安全稳定运行[22]。

$$\|r\| = \|z_a - H\hat{x}_a\| = \|z - H\hat{x}\| \quad (17)$$

若进一步将虚假数据误差考虑在内，攻击前后的量测残差将不再相等[18]，如下式所示：

$$\|z_a - H\hat{x}_a\| = \|z - H\hat{x} + a - Hc\|$$
$$\leq \|z - H\hat{x}\| + \|a - Hc\| \quad (18)$$

然而，如果量测数据的残差值小于不良数据的检测阈值，FDI 攻击仍然可以成功实施。在基于残差的不良数据检测中，将正常情况下最大估计偏差值增加一定的冗余度，形成检测阈值 $B_J$。其检测过程如下式所示：

$$\|z - H\hat{x}\| \leq \|z_a - H\hat{x}_a\| \leq B_J \quad (19)$$

即如果 $z_a$ 满足式（19），FDI 攻击将会被成功引入发电机动态状态估计中。

## 3 容积卡尔曼滤波与抗差容积卡尔曼滤波

### 3.1 容积卡尔曼滤波

CKF算法采用球面-径向准则生成一组等权值的容积点，进而解决了发电机动态状态估计中的非线性滤波问题，其具体分为两部分：

1）预报步。CKF 根据球面-径向准则产生一组等权值的状态量容积点，之后利用状态方程获得下一时刻状态量的预报值，并计算状态量预报误差方差矩阵。其具体步骤如下：

利用 $k$ 时刻状态量的估计误差方差矩阵得到其平方根矩阵：

$$P_{k/k} = S_{k/k}S_{k/k}^T \quad (20)$$

生成状态量的容积点：

$$X_{i,k/k} = S_{k/k}\zeta_i + \hat{x}_{k/k} \quad (21)$$

式中 $\xi_i$ 为容积点权值矩阵 $\sqrt{n}[I_n \ -I_n]$ 的第 $i$ 列，$I_n$ 表示单位阵，$n$ 表示状态量个数。$\hat{x}_{k/k}$ 表示 $k+1$ 时刻状态量估计值。

经状态方程计算得到状态量容积点预报值：

$$X_{i,k/k}^* = f(X_{i,k/k}, u_k) \quad (22)$$

对上式结果进行加权求和计算，进一步得到状态量预报值：

$$\hat{x}_{k+1/k} = \frac{1}{2n}\sum_{i=1}^{2n} X_{i,k+1/k}^* \quad (23)$$

计算获取状态量预报误差方差矩阵：

$$P_{k+1/k} = \frac{1}{2n}\sum_{i=1}^{2n} X_{i,k+1/k}^* X_{i,k+1/k}^{*T} - \hat{x}_{k+1/k}\hat{x}_{k+1/k}^T + Q_k \quad (24)$$

2）滤波步。通过量测量完成对状态量预报值的修正，得到更为准确的估计值。其具体过程如下：

计算状态量预报误差方差矩阵的平方根矩阵：

$$P_{k+1/k} = S_{k+1/k}S_{k+1/k}^T \quad (25)$$

计算状态量预报值的容积点：

$$X_{i,k+1/k} = S_{k+1/k}\zeta_i + \hat{x}_{k+1/k} \quad (26)$$

通过量测方程计算量测预报值容积点：

$$Z_{i,k+1/k} = h(X_{i,k+1/k}, u_k) \quad (27)$$

对上式结果进行加权求和计算，进一步得到量测量预报值：

$$\hat{z}_{k+1/k} = \frac{1}{2n}\sum_{i=1}^{2n} Z_{i,k+1/k} \quad (28)$$

建立量测量误差方差矩阵：

$$P_{zz,k+1/k} = \frac{1}{2n}\sum_{i=1}^{2n} Z_{i,k+1/k}Z_{i,k+1/k}^T - \hat{z}_{k+1/k}\hat{z}_{k+1/k}^T + R_{k+1} \quad (29)$$

计算交叉误差方差矩阵：

$$P_{xz,k+1/k} = \frac{1}{2n}\sum_{i=1}^{2n} X_{i,k+1/k}Z_{i,k+1/k}^T - \hat{x}_{k+1/k}\hat{z}_{k+1/k}^T \quad (30)$$

计算滤波增益：

$$W_{k+1} = P_{xz,k+1/k}P_{zz,k+1/k}^{-1} \quad (31)$$

计算 $k+1$ 时刻状态量估计值：

$$\hat{x}_{k+1/k+1} = \hat{x}_{k+1/k} + W_{k+1}(z_{k+1} - \hat{z}_{k+1/k}) \quad (32)$$

计算下一时刻状态量估计误差方差矩阵：

$$P_{k+1/k} = P_{k+1/k} - W_{k+1}P_{zz,k+1/k}W_{k+1}^T \quad (33)$$

### 3.2 抗差容积卡尔曼滤波

RCKF 算法是抗差估计理论中的中位数估计方法和 CKF 算法的融合。在抗差估计理论中，M 估计是经典极大似然估计的推广，又称为广义似然估计，在电力系统中应用极为广泛。对量测方程：

$$Z = Ax + G \quad (34)$$

式中 $A$ 为 $m×n$ 阶系数矩阵，$x$ 是状态向量，$Z$ 是量测误差向量，$G$ 是 $Z$ 的残差向量。

M 估计的准则函数是：

$$\sum_{i=1}^{m}\rho(g_i) = \min \quad (35)$$

其中，$\rho(\cdot)$ 为适当选择的凸函数；$g_i$ 为量测残差。

作为 M 估计中的一种估计方法，中位数估计方法通过设置残差的中位数来滤除量测量中的粗差，与其他的传统最小二乘抗差方法相比能够避免更多的量测不良数据对估计结果的影响，具有良好的抗差性能。在中位数估计中，$\rho(\cdot)$ 函数和等价权因子 $p(\cdot)$ 分别如下：

$$\rho(g_i) = |g_i| \quad (36)$$

$$p(g_i) = \frac{1}{|g_i|} \quad (37)$$

其中，$g_i = (z_i - a_ix)/\text{median}(z_i - a_ix)$，$\text{median}(\cdot)$ 表示对 $(\cdot)$ 取中位数。

经上述方法，首先根据中位数估计方法确定预测残差中误差，并计算标准化残差，然后基于标准化残差绝对值，通过两段函数获得量测噪声的抗差协方差矩阵，最后用其替换原来的量测误差方差矩阵，因此，在算法计算过程中，量测量误差方差矩阵被不断修正，进而形成 RCKF 算法。其具体过程如下：

计算量测量的预测残差向量：

$$r_k = z_{k+1} - \hat{z}_{k+1/k} \quad (38)$$

计算预测残差中误差：
$$\sigma r_k = \text{median}(|r_k|) \quad (39)$$

计算标准化残差：
$$r_k^{'} = r_k / \sigma r_k \quad (40)$$

计算抗差协方差矩阵：
$$R_{k+1,ii}^{'} = \begin{cases} R_{k+1,ii}, |r_{i,k}^{'}| \leq C_i \\ \dfrac{C_i}{|r_{i,k}^{'}|} R_{k+1,ii}, |r_{i,k}^{'}| > C_i \end{cases} \quad (41)$$

上式中，$R_{k+1,ii}^{'}$ 和 $R_{k+1,ii}$ 分别为 $R_{k+1}^{'}$ 和 $R_{k+1}$ 的第 $i$ 个对角元素，非对角元素为零，$C_i$ 为先验阈值（即检测量测数据异常的阈值），为一常数。由于 RCKF 是以 CKF 为基础的算法，因此需要通过 CKF 算法计算的标准化残差值来确定 $C_i$。具体步骤如下：（1）在确定的系统中预先设置线路或母线故障，对于给定的发电机进行基于 CKF 算法的动态状态估计，通过式（32）计算得到量测量的标准化残差随时间的变化曲线；（2）根据曲线的分布情况确定阈值：如果曲线收敛，则取其收敛值作为 RCKF 算法中的先验阈值；否则，取介于最大值和最小值的中间值作为先验阈值。

更新量测量误差方差矩阵：
$$P_{zz,k+1/k} = \frac{1}{2n}\sum_{i=1}^{2n} Z_{i,k+1/k} Z_{i,k+1/k}^T - \hat{z}_{k+1/k}\hat{z}_{k+1/k}^T + R_{k+1}^{'} \quad (42)$$

将式（44）替换式（29），进而形成 RCKF 算法。

## 4 估计流程

FDI 攻击场景下，本文中发电机动态状态估计的步骤如下：

（1）建立发电机状态方程，将 $k$ 时刻状态量的估计值和由 PMU 实时测得的发电机出口电压幅值及相角 $(U,\varphi)$ 带入发电机状态方程，进行预报步的计算，从而得到 $k+1$ 时刻状态量的预报值。

（2）建立发电机量测方程，根据量测方程求出联系量测量和状态量之间的雅克比矩阵。分别建立服从不同标准差高斯分布的状态量误差向量。将两者相乘形成攻击向量，让其施加在由 PMU 测得的量测数据中，并进一步对量测数据进行不良数据检测。

（3）在发电机量测方程和更新后量测值的基础上，通过 CKF 算法或 RCKF 算法的滤波阶段，进一步得到该攻击场景下 $k+1$ 时刻状态量的估计值。

（4）判断仿真时间是否结束，如果结束则停止估计过程，否则将此估计值代入步骤（1）进行下一时刻的状态估计。

FDI 攻击场景下，本文中发电机动态状态估计的流程如下图所示：

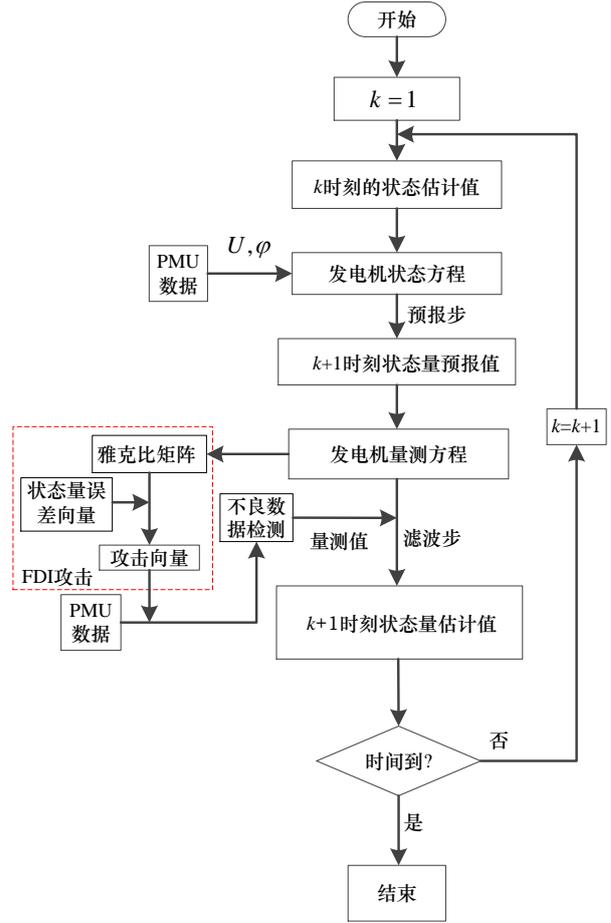

图 1 状态估计流程图
Figure 1. Flowchart of the state estimation

## 5 仿真分析

### 5.1 仿真系统设置

在 MATLAB 环境下，采用 Power System Toolbox（PST）为仿真工具。所有仿真均在配置为 Intel Core i5-4590 3.3 GHz 处理器、8 GB 内存的 PC 机上实现。仿真基准频率为 50Hz，采样步长为 0.02s；PMU 量测误差服从如下正态分布：均值为零，发电机功角标准差是 2°，角速度标准差是 0.1%，出口电压的相角和幅值标准差是 0.1°和 0.1%[1]。

结合文献[23]中基于 FDI 攻击下状态量的误差向量 $c$ 的选取结果，为了进一步研究更大程度的攻击对估计结果的影响，因此，在本文的仿真分析中，根据 FDI 攻击程度由弱到强共设计了三种攻击情形，如表 1 所示。

表 1 虚假数据注入攻击情形
Table 1 False data injection attack situation

| 攻击情形 | 状态量的误差向量 |
| --- | --- |
| Case 1 | $c \sim N(0,\sigma^2), \sigma = 0.01$ |
| Case 2 | $c \sim N(0,\sigma^2), \sigma = 0.1$ |
| Case 3 | $c \sim N(0,\sigma^2), \sigma = 1$ |

在表 1 中，Case 1、Case 2 和 Case 3 分别表示在此三种攻击情形下，状态量的误差向量分别服从均值为零，标准差为 0.01、0.1 和 1 的高斯分布。

### 5.2 评估指标

建立发电机动态状态估计的量化评估指标 $\tau_1$ 和总估计误差方差 $\tau_2$，具体计算公式如下[1]：

$$\tau_1 = \sqrt{\frac{1}{N}\sum_{i=1}^{N}(\frac{\hat{x}_i - x_{it}}{x_{iz}})^2} \qquad (43)$$

$$\tau_2 = \sqrt{\frac{\sum_{i=1}^{N}(\hat{x}_i - x_{it})^2}{\sum_{i=1}^{N}(x_{iz} - x_{it})^2}} \qquad (44)$$

式中，$\hat{x}_i$ 和 $x_{it}$ 分别为估计值和真实值，$x_{iz}$ 为量测值，$N$ 为采样点数，评估指标 $\tau_1$ 可衡量基于同一估计方法在不同攻击场景下的滤波性能；$\tau_2$ 可评价在同一信息攻击下不同估计方法的滤波性能。

### 5.3 IEEE 9 节点系统

将所提算法用于 IEEE 9 节点系统进行仿真，仿真设置如下：在 $t$=1.2 s 时在节点 5 设置三相金属性短路故障，随后故障被断路器切除，仿真持续时间 20 s。经过多次仿真测试，设置 $B_J = 2.0$。在此情况下，不良数据检测能够检测到不良数据但是检测不到注入的虚假数据。

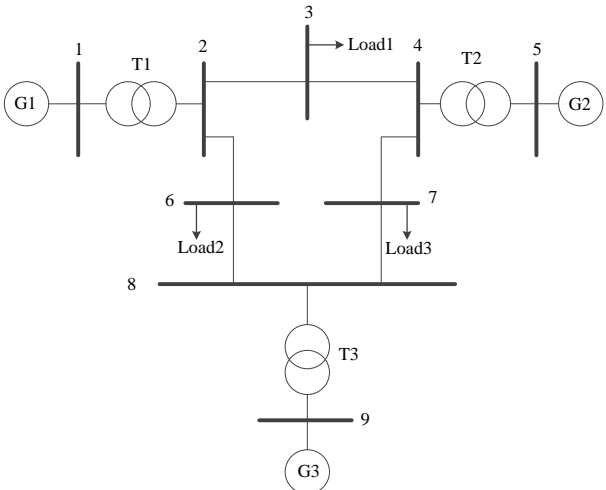

图 2  IEEE 9 节点系统
Figure 2.  IEEE 9-bus system

以发电机 1 为例，下图所示为基于 CKF 算法的 3 个量测量标准化残差的变化情况。

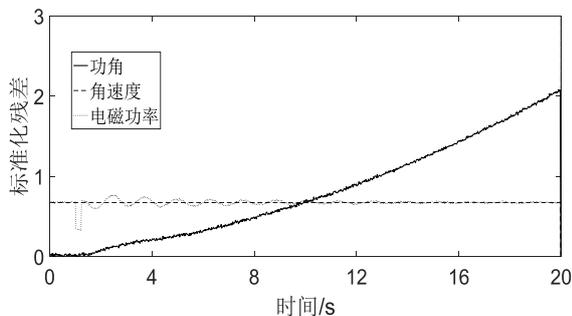

图 3  基于 CKF 算法的标准化残差
Figure 3.  Standardized residuals based on CKF algorithm

由上图可得，由于量测异常值的存在，CKF 计算下的标准化残差变化不一，其中功角的标准化残差逐渐增大，角速度的标准化残差比较稳定，电磁功率的标准化残差在前期虽存在跳变，但后期主体趋势逐渐稳定，故对 3 个先验阈值分别取值 1、0.7 和 0.7。

以发电机 1 为例，分别在上述三种攻击情形下，进行基于 RCKF 和 CKF 的发电机动态状态估计，得到以下发电机功角和角速度的仿真图形。

**（1）无攻击情形**

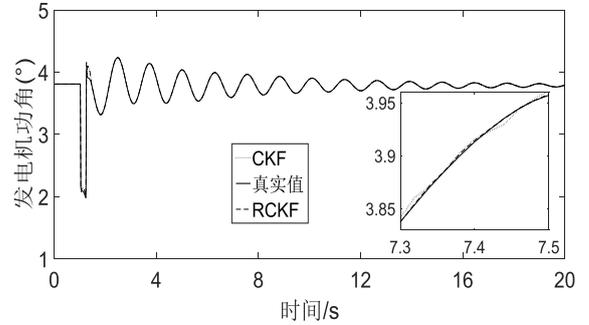

图 4  无攻击情形下的发电机功角
Figure 4.  Generator's power angle without attack

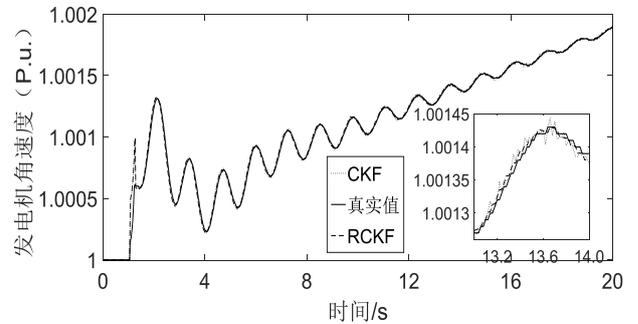

图 5  无攻击情形下的发电机角速度
Figure 5.  Generator's angular velocity without attack

**（2）Case 1-情形 1**

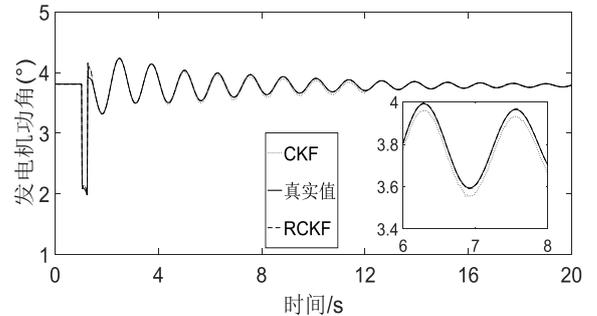

图 6  情形 1 下的发电机功角
Figure 6.  Generator's power angle under case 1

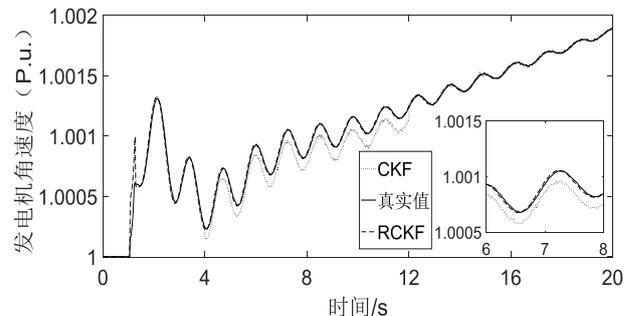

图 7  情形 1 下的发电机角速度
Figure 7.  Generator's angular velocity under case 1

**（3）Case 2-情形 2**

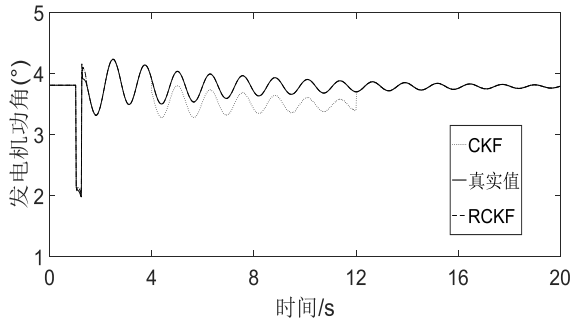

图 8  情形 2 下的发电机功角
Figure 8.  Generator's power angle under case 2

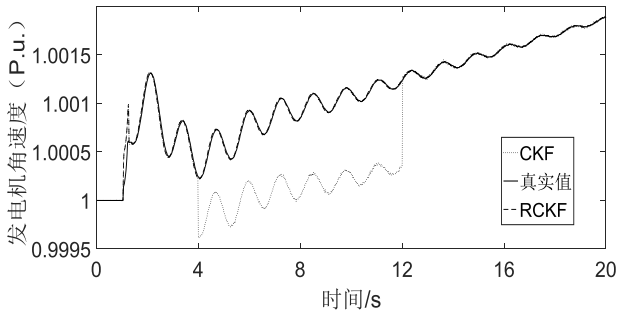

图 9  情形 2 下的发电机角速度
Figure 9.  Generator's angular velocity under case 2

**(4) Case 3-情形 3**

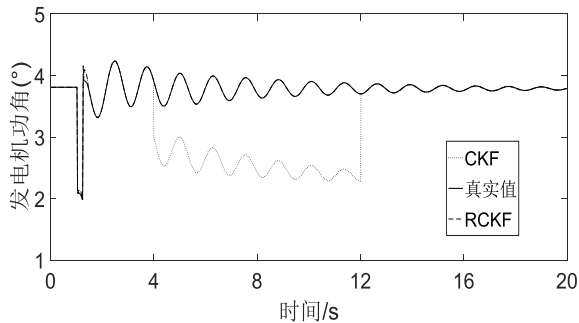

图 10  情形 3 下的发电机功角
Figure 10.  Generator's power angle under case 3

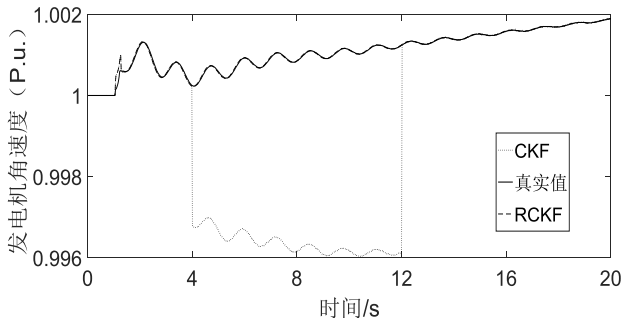

图 11  情形 3 下的发电机角速度
Figure 11.  Generator's angular velocity under case 3

图 4 和图 5 显示了无攻击情形下分别基于 CKF 和 RCKF 的发电机功角和角速度的估计结果；图 6, 8, 10 分别显示了在三种 FDI 攻击场景下，发电机功角的估计结果；图 7, 9, 11 分别显示了在三种 FDI 攻击场景下，发电机角速度的估计结果。可以看出在第 4 秒至第 12 秒的攻击时段内，与无攻击情形相比，随着攻击强度的增加，CKF 的估计值逐渐偏离真实值，从而验证了所提 FDI 攻击对基于 CKF 算法的发电机动态状态估计的有效性。随着攻击情形的变化，RCKF 算法对攻击具有一定的抵抗能力，其估计值未明显偏移。

**（5）状态估计指标对比**

发电机 1 的状态估计指标如表 2 所示。

表 2  发电机 1 的动态状态估计指标
Table 2 Dynamic state estimation index of generator 1

| 攻击情形 | 指标 | 参数 | CKF | RCKF |
|---|---|---|---|---|
| 无攻击 | $\tau_1$ | $\delta$ | 7.5080e-04 | 4.8887e-04 |
|  |  | $\omega$ | 2.7551e-04 | 2.5947e-04 |
|  | $\tau_2$ | $\delta$ | 2.1000e-03 | 1.2000e-03 |
|  |  | $\omega$ | 1.7776e-05 | 1.6294e-05 |
| Case 1 | $\tau_1$ | $\delta$ | 7.7000e-03 | 4.8893e-04 |
|  |  | $\omega$ | 2.0000e-03 | 2.6016e-04 |
|  | $\tau_2$ | $\delta$ | 1.7100e-02 | 1.3000e-03 |
|  |  | $\omega$ | 9.1503e-05 | 1.6410e-05 |
| Case 2 | $\tau_1$ | $\delta$ | 6.0800e-02 | 4.8931e-04 |
|  |  | $\omega$ | 1.4100e-02 | 2.6037e-04 |
|  | $\tau_2$ | $\delta$ | 4.0900e-02 | 3.4311e-04 |
|  |  | $\omega$ | 8.0718e-04 | 1.7780e-05 |
| Case 3 | $\tau_1$ | $\delta$ | 7.6320e-01 | 4.8948e-04 |
|  |  | $\omega$ | 1.8470e-01 | 2.6048e-04 |
|  | $\tau_2$ | $\delta$ | 4.7600e-02 | 2.8935e-05 |
|  |  | $\omega$ | 5.2000e-03 | 7.6539e-06 |

由表 2 可知：在无攻击和三种不同攻击情形下，基于 CKF 和 RCKF 的各指标值均有所变化。（1）基于状态指标值 $\tau_1$，在分别面对无攻击和三种逐渐增强的攻击时，CKF 的功角指标值和角速度指标值逐渐增加且增幅很大，与此同时，RCKF 的功角指标值和角速度指标值也逐渐增加，说明 RCKF 虽具有抗差能力，但在面对逐渐增强的 FDI 攻击时，其状态估计精度仍有所下降，即验证了所提 FDI 攻击对基于 CKF 和 RCKF 的发电机动态状态估计的有效性。（2）基于状态指标值 $\tau_2$，在面对无攻击和三种攻击情形时，RCKF 的滤波精度均大于 CKF。因为在无攻击情形下，结合了中位数估计方法的 RCKF 有效滤除了量测噪音，因此其抗噪性能优于 CKF；在面对三种逐渐增强的攻击时，由于具有抗差能力，RCKF 在一定程度上削弱了攻击向量对其估计精度的影响，因而 RCKF 的滤波能力优于 CKF。

## 5.4 新英格兰 16 机 68 节点系统

该系统为状态估计领域的经典测试算例[16]，由 16 台同步发电机，68 条母线，86 条传输线路组成。$t=1$ s 时在节点 6 设置三相金属性短路故障，1.05 s 时切除节点 1 的近端故障，1.1 s 时切除节点 54 的远端故障，仿真持续时间 10s。与在 IEEE 9 节点系统中对 $B_J$ 的设置方法一样，在新英格兰 16 机 68 节点系统中设置 $B_J = 1.5$。

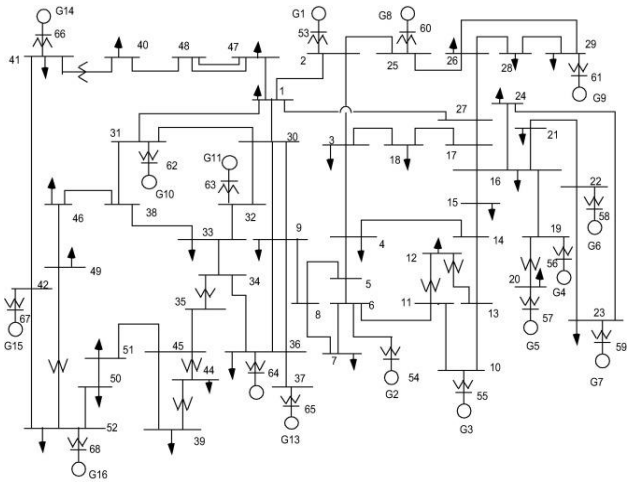

图 12　新英格兰 16 机 68 节点系统
Figure 12. New England 16-machine-68-bus system

采用 CKF 算法时，量测量的残差变化情况如下图所示。

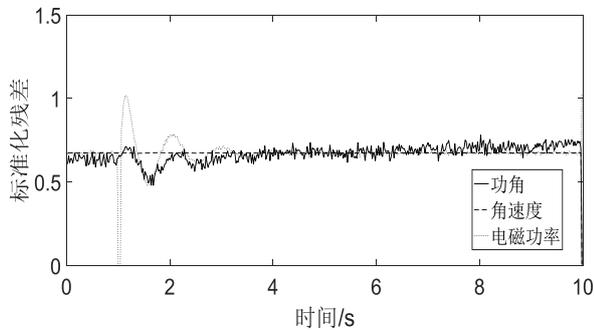

图 13　基于 CKF 算法的标准化残差
Figure 13. Standardized residuals based on CKF algorithm

由上图可知，由于 PMU 量测异常值的存在，CKF 算法下 3 个量测量的标准化残差仍然存在大量的跳变，但主体趋势稳定，故在此系统下对 3 个先验阈值均取 0.67。

与 IEEE 9 节点系统的攻击情形设置相同，以发电机 1 为例，分别在三种攻击情形下，基于两种算法进行仿真，得到如下仿真结果。

（1）无攻击情形

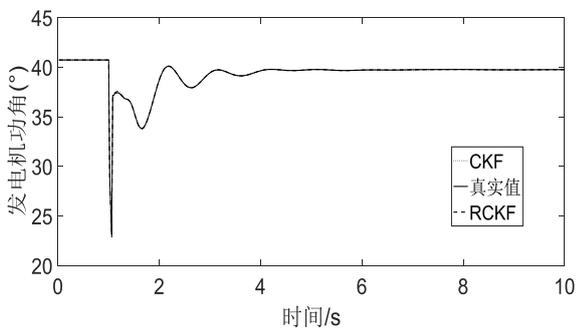

图 14　无攻击情形下的发电机功角
Figure 14. Generator's power angle without attack

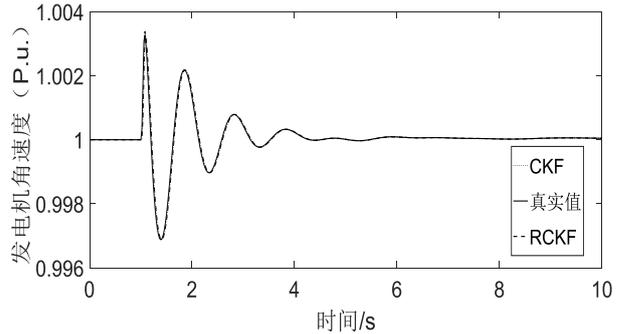

图 15　无攻击情形下的发电机角速度
Figure 15. Generator's angular velocity without attack

（2）Case 1-情形 1

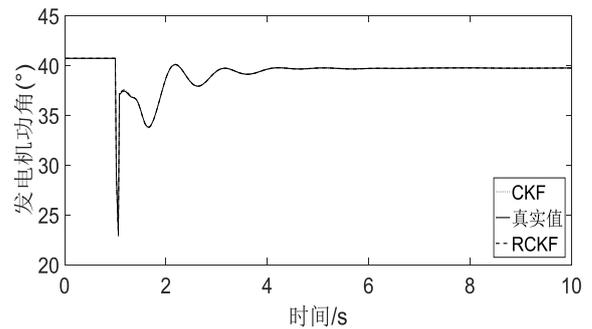

图 16　情形 1 下的发电机功角
Figure 16. Generator's power angle under case 1

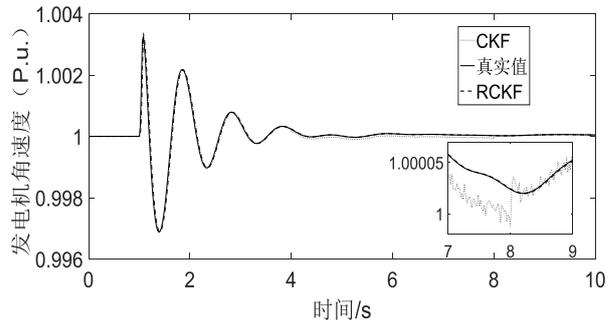

图 17　情形 1 下的发电机角速度
Figure 17. Generator's angular velocity under case 1

（3）Case 2-情形 2

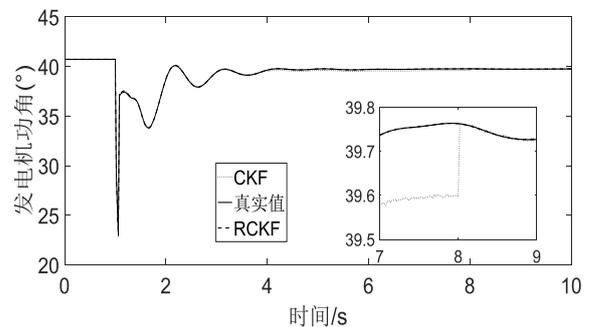

图 18　情形 2 下的发电机功角
Figure 18. Generator's power angle under case 2

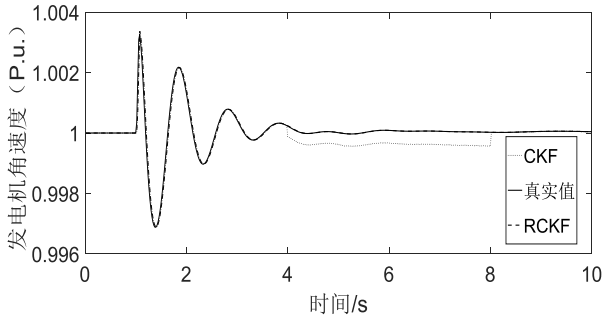

图 19 情形 2 下的发电机角速度
Figure 19. Generator's angular velocity under case 2

**（4）Case 3-情形 3**

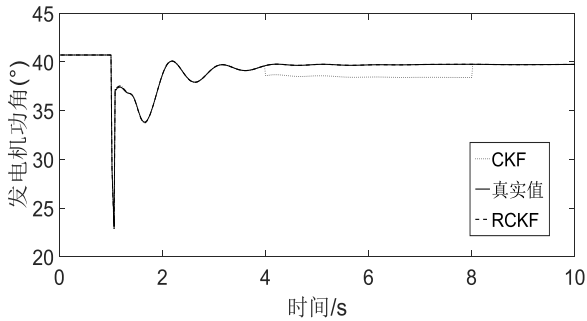

图 20 情形 3 下的发电机功角
Figure 20. Generator's power angle under case 3

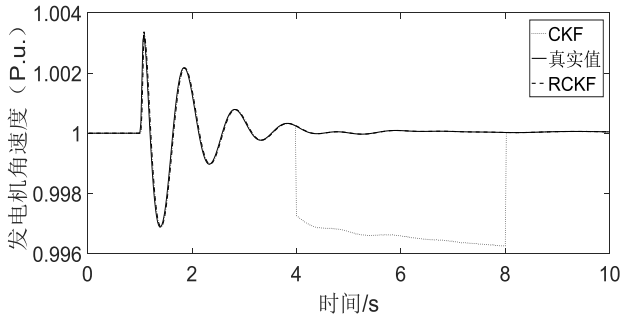

图 21 情形 3 下的发电机角速度
Figure 21. Generator's angular velocity under case 3

图 14 和图 15 分别显示了无攻击下的发电机功角和角速度的估计结果；图 16, 18, 20 分别显示了在三种攻击情形下发电机功角的估计结果；图 17, 19, 21 分别显示了三种攻击情形下发电机角速度的估计结果。可以看出在第 4~8 秒的攻击时间内，面对无攻击情形和三种逐渐增强的攻击时，CKF 的估计值越来越偏离真实值，因此验证了所提 FDI 攻击在较大规模系统中对基于 CKF 的发电机动态状态估计的有效性。与此同时，随着攻击情形的变化，RCKF 算法对攻击具有一定的抵抗能力，估计值未出现明显偏移。

**（5）状态估计指标对比**

发电机 1 的状态估计指标如表 3 所示。

表 3 发电机 1 的动态状态估计指标
Table 3 Dynamic state estimation indicators of generator 1

| 攻击情形 | 指标 | 参数 | CKF | RCKF |
|---|---|---|---|---|
| 无攻击 | $\tau_1$ | $\delta$ | 4.4829e-04 | 1.3025e-04 |
| | | $\omega$ | 5.9986e-05 | 1.4660e-05 |
| | $\tau_2$ | $\delta$ | 1.2000e-03 | 4.4290e-04 |
| | | $\omega$ | 8.3648e-05 | 5.4481e-06 |
| Case 1 | $\tau_1$ | $\delta$ | 1.8000e-03 | 1.3038e-04 |
| | | $\omega$ | 3.0482e-04 | 1.4682e-05 |
| | $\tau_2$ | $\delta$ | 3.9000e-03 | 4.5320e-04 |
| | | $\omega$ | 3.6726e-04 | 2.0538e-05 |
| Case 2 | $\tau_1$ | $\delta$ | 2.7300e-02 | 1.3048e-04 |
| | | $\omega$ | 4.8000e-03 | 1.4689e-05 |
| | $\tau_2$ | $\delta$ | 1.8900e-02 | 1.3919e-04 |
| | | $\omega$ | 2.4000e-03 | 8.1712e-06 |
| Case 3 | $\tau_1$ | $\delta$ | 1.3730e-01 | 1.3079e-04 |
| | | $\omega$ | 2.3900e-02 | 1.4697e-05 |
| | $\tau_2$ | $\delta$ | 2.2200e-02 | 3.4607e-05 |
| | | $\omega$ | 3.3000e-03 | 2.5264e-06 |

由表 3 可知，面对不同的攻击情形时，CKF 和 RCKF 的指标值均有所变化。（1）对于指标 $\tau_1$，随着攻击程度的逐渐增强，基于 RCKF 的发电机功角和角速度的指标值均逐渐递增，而基于 CKF 的发电机功角和角速度的指标值大幅增加，表明了基于 CKF 和 RCKF 的估计结果精度均受到 FDI 攻击的影响，从而验证了在新英格兰 16 机 68 节点系统中，所提 FDI 攻击对基于这两种算法的发电机动态状态估计的有效性。（2）对于指标 $\tau_2$，在面对无攻击和三种攻击情形时，RCKF 的滤波精度均大于 CKF。因为在无攻击情形下，RCKF 滤除了量测量中存在的噪音；在面对三种逐渐增强的攻击时，由于 RCKF 具有抗差能力，其一定程度上削弱了量测量中攻击向量对估计精度的影响，所以 RCKF 的滤波性能优于 CKF。

对比表 2 和表 3，可得：（1）就指标 $\tau_1$ 而言，在两个测试算例中，面对不同攻击情形，RCKF 的指标值和 CKF 的指标值均有不同程度的增加，验证了本文所提的 FDI 攻击对基于这两种算法的发电机动态状态估计的有效性；（2）就指标 $\tau_2$ 而言，两种测试系统的仿真结果表明，基于 RCKF 的指标值均小于 CKF，证实了 RCKF 较 CKF 具有良好的滤波性能。

### 5.5 计算效率比较

以上测试中，两种算法的计算时间如表 4 所示。

表 4 计算时间
Table 4 Calculating time

| 测试系统 | 攻击情形 | CKF（ms） | RCKF（ms） |
|---|---|---|---|
| IEEE 9 节点系统 | 无攻击 | 1.24 | 3.06 |
| | Case 1 | 1.36 | 3.07 |
| | Case 2 | 1.37 | 3.09 |
| | Case 3 | 1.42 | 3.13 |
| 新英格兰 16 机 68 节点系统 | 无攻击 | 1.54 | 3.48 |
| | Case 1 | 1.57 | 3.53 |
| | Case 2 | 1.61 | 3.55 |
| | Case 3 | 1.64 | 3.56 |

由表 4 可知，因为需要经过抗差计算处理，因此算法的计算时间略长于 CKF 算法，但仍远小于 PMU 采样间隔（20ms）。故而，RCKF 仍能满足发电机动态状态估计对实时性的应用需求。

## 6 结论

本文提出了一种发电机动态状态估计的虚假数据注入攻击方法。该方法首先建立了 FDI 攻击模型，并将攻击向量施加在量测量中，然后根据攻击程度分别设定三种攻击情形，最后，在此基础上分别进行基于 CKF 和 RCKF 的发电机动态状态估计仿真。两个仿真

系统的测试结果表明：（1）所提 FDI 攻击均不同程度地恶化了基于两种算法的发电机动态状态估计结果，验证了所提方法的有效性；（2）在无攻击情形和所提 FDI 攻击场景下，RCKF 的滤波性能优于 CKF。

文中仅考虑了虚假数据注入攻击，下一步将研究其他类型信息攻击场景下的发电机状态估计。文中不良数据检测阈值 $B_t$ 是基于经验、经过多次仿真测试来确定的，后续研究中将进一步设计一种通用的阈值选取方法。此外，将所提方法扩展至整个电力系统的状态估计也是值得研究的课题[26-29]。

## 参考文献

作者简介

李扬 男, 1980 年生, 博士, 副教授, 研究方向为电力系统运行分析与控制。E-mail: liyang@neepu.edu.cn (通信作者)

李智 男, 1994 年生, 硕士研究生, 研究方向为电力系统运行分析与控制。E-mail: 1782798766 @qq.com